\documentclass[twocolumn,showpacs,preprintnumbers,amsmath,amssymb,pra,epsf]{revtex4}
\usepackage{graphicx}
\begin{document}
\title{Correlation measurement of squeezed light}

\author{Leonid A. Krivitsky$^{1,2,3}$, Ulrik L. Andersen$^1$, Ruifang Dong$^2$, Alexander Huck$^1$, Christoffer Wittmann$^2$, and Gerd Leuchs$^2$}

\affiliation{
$^1$ Technical University of Denmark, Department of Physics,
Building 309, 2800 Lyngby, Denmark\\
$^2$ Max Planck Institute for the Science of Light, G\"unther-Scharowsky-Str.
1 / Bau 24, 91058 Erlangen, Germany\\
$^3$ Data Storage Institute, Agency for Science, Technology and Research (A-Star), 117608 Singapore
$^4$ Institute of Optics, Information, and Photonics, Friedrich-Alexander-University Erlangen-Nuremberg, Staudtstraße. 7/B2, 91058 Erlangen, Germany}

\begin{abstract}
\begin{center}\parbox{14.5cm}
{We study the implementation of a correlation measurement technique
for the characterization of squeezed light which is nearly free of electronic noise. With two different sources of squeezed light, we show that the sign of the covariance coefficient, revealed from the time resolved correlation data, is witnessing the presence of squeezing in the system. Furthermore, we estimate the degree of squeezing using the correlation method and compare it to the standard homodyne measurement scheme. We show that the role of electronic detector noise is minimized using the correlation approach as opposed to homodyning where it often becomes a crucial issue.}
\end{center}
\end{abstract}
\pacs{42.50.Dv, 03.67.Hk, 42.62.Eh}
\maketitle \narrowtext
\vspace{-10mm}

\section{Introduction}
Since the pioneering experiments of Hanbury-Brown and Twiss ~\cite{twiss} (HBT) on the realization of the stellar intensity interferometer, the correlation measurement technique has proven its great potential in modern quantum optics. A direct observation of the second order intensity correlation function reveals many nontrivial properties of the photon statistics~\cite{mandelbook}. For example, the observation of the strong photon bunching is common in experiments where correlated photon pairs are generated in various nonlinear optical processes. In practice, such a technique is involved in many fundamental and practical tasks such as tests of the foundation of quantum mechanics~\cite{chuang}, quantum cryptography~\cite{gisin}, quantum metrology~\cite{alan} and many others.

In contrast, in experiments with squeezed light, normally associated with the detection of a continuous degree of freedom, a different measurement technique is traditionally implemented. The conventional method applied in this case is balanced homodyne detection (HD) where the state under investigation (often a squeezed state) is mixed on a symmetric beamsplitter with a local oscillator (LO) and the difference
photocurrent of two analogue photodetectors is recorded~\cite{review} (see also Fig.1). By scanning the phase of the LO the statistics of the field's quadratures are probed which subsequently allows for the full tomographic characterization of the state~\cite{raymer}.
 
Recently, there has been a trend to combine the measurement techniques of the discrete and continuous variable domains.  In ref.~\cite{lvovsky, bellini} HD techniques were applied to the characterization of a single photon Fock state as an alternative to the traditional correlation method. Moreover the characterization of a squeezed state through measurements of the second-order correlation function - commonly used in the discrete variable quantum optics - was theoretically proposed in ~\cite{hom} and experimentally demonstrated using single photon detectors~\cite{kono} and homodyne detectors~\cite{pklam}.

In the appendix of ref.~\cite{leuchs}, the correlation measurement technique of squeezed light was analyzed and it was found that it provides an electronic-noise free approach for estimating the degree of squeezing. This electronic-noise free estimation method was experimentally implemented in ref.~\cite{kriv} for a bright squeezed light beam generated in a fiber exploiting the optical Kerr effect. In the present paper we extend the experimental analysis to include vacuum squeezed light produced by an optical parametric oscillator. The utilization of squeezed vacuum yields a more thorough analysis of the effect of electronic noise on the measurement of squeezing.

\section{Theory}

Let us consider a HD setup as shown in Fig.1: The signal field under investigation is mixed with a LO on a 50/50 beam splitter. The two outputs are measured with analogue detectors and the difference of the photocurrents of the two detectors is produced. The theoretical treatment is straightforward and follows the standard formalism~\cite{bachorbook}. The field operator can be written as $a=\alpha+\delta{a}$, where $\alpha$
represents the classical bright component and $\delta{a}$ is an
operator with zero mean value, describing the quantum
uncertainty of the field amplitude. We consider the input state to be a \emph{squeezed vacuum} state. Assuming unity detector efficiency
we calculate the photocurrents
of the two detectors in each port of the beamsplitter:
\begin{equation}
i_{1,2}=1/2\alpha^2_{LO}+\alpha_{LO}(\delta{X_{LO}}\pm\delta{X_{\phi}})+\delta{i_{el1,2}}\\
\end{equation}
where $\alpha_{LO}$ and $\phi$ is the mean field amplitude and the phase of the LO, respectively, $\delta
X_\phi\equiv\cos{\phi}\delta X+\sin{\phi}\delta Y$ is the
fluctuation of the rotated quadrature of the input beam,
$\delta{X_{LO}}$ is the amplitude quadrature of the LO and
$\delta{i_{el1,2}}$ is a stochastic variable accounting for the electronic noise (EN). To complete the homodyne setup the 
difference photocurrent is recorded and its variance is given by
\begin{equation}
\langle(i_{1}-i_{2})^2\rangle=4\alpha^2_{LO}\langle\delta{X^2_{\phi}}\rangle+\langle\delta{i^2_{el1}}\rangle+\langle\delta{i^2_{el2}}\rangle.
\label{homodyne}
\end{equation}
which is clearly seen to be measuring the quadrature fluctuations of the signal, $\langle\delta X_\phi^2\rangle$. However, it is also clear that the contribution from the EN always affects the measurement: The ENs are additive and their presence might dominate the contribution from the signal thus potentially masking the quantum features. The EN relative to the signal noise can be made very small partly through a clever design of the amplifying circuit of the detectors and partly by using a very strong LO which "amplifies" the signal contribution but leaving the EN unchanged. This is however not always possible and the performance of the HD will be ultimately limited by the EN. 

For the estimation of squeezing, a careful calibration of the shot noise level (SNL) must be carried out. The SNL calibration is however also affected by the EN and it is normally performed by blocking the input signal, thus replacing it by vacuum. It yields the variance
\begin{equation}
\langle(i_{1}-i_{2})^2\rangle_{"SNL"}=4\alpha^2_{LO}\langle\delta{X^2_{vac}}\rangle+\langle\delta{i^2_{el1}}\rangle+\langle\delta{i^2_{el2}}\rangle.
\label{SNL}
\end{equation}
which is clearly a false estimation of the SNL due to the EN contribution. To reliably estimate the SNL, an independent measurement of the EN level must be executed (by blocking all input beams to the detectors). At very high LO powers, where the contribution from the EN is relatively low, it can be subsequently subtracted from the measurement outcome in (\ref{SNL}) thus yielding 
\begin{equation}
V_{SNL}=4\alpha^2_{LO}.
\label{SNL2}
\end{equation} 
And the SNL for all powers can then be found by extrapolation using the fact that the SNL scales linearly with the power of the LO. We note that such a calibration can be performed with any bright light beam provided that the homodyne detector is well balanced or the light beam is shot noise limited. There is no requirements for spatio-temporal matching with the signal beam (which is just the vacuum state).    
 
A traditional squeezing measurement is therefore carried out by recording the variance of the difference current (using a spectrum analyzer) and subsequently compare this to the shot noise level:
\begin{eqnarray}
S &=&\frac{\langle(i_{1}-i_{2})^2\rangle_{\delta X_{\phi}}}{V_{SNL}}\nonumber
\\
&=&\frac{1}{4\alpha_{LO}^2}\left(4\alpha_{LO}^2\langle\delta X_\phi^2\rangle+\langle\delta{i^2_{el1}}\rangle+\langle\delta{i^2_{el2}}\rangle\right)
\label{squeezing}
\end{eqnarray}
As clearly seen from this expression, the measurement of the degree of squeezing is hampered by the electronic noise of the detectors; only if $\langle\delta{i^2_{el1}}\rangle+\langle\delta{i^2_{el2}}\rangle<<4\alpha_{LO}^2\langle\delta X_\phi^2\rangle$, $S\rightarrow\langle\delta X_\phi^2\rangle$.   

\begin{figure}
\includegraphics[width=8cm]{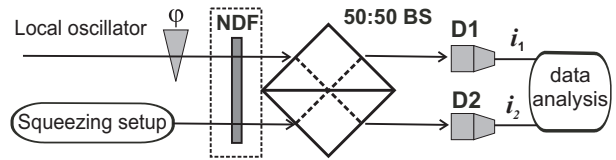}
\caption{Schematic of the setup. The beam from the squeezing setup (OPA or Kerr based squeezing setup, see text) is interfering with the LO on a 50/50 beamsplitter \textbf{BS}, with a relative phase of $\phi$. \textbf{D1} and \textbf{D2} are the analogue photodetectors. The photocurrents of the detectors $i_1$ and $i_2$ are analyzed by the acquisition system.}
\label{fig1}
\end{figure}

Instead of constructing the difference current, it is interesting to consider the covariance of the two detector outcomes:
\begin{eqnarray}
\hbox{cov}(i_{1},i_{2})\equiv\langle{i_{1}i_{2}}\rangle-\langle{i_{1}}\rangle\langle{i_{2}}\rangle=\langle{a ^\dagger_{1}a_{1}a ^\dagger_{2}a_{2}}\rangle-
\nonumber\\
\langle a ^\dagger_{1}a_{1}\rangle\langle a ^\dagger_{2}a_{2}\rangle=\alpha^2_{LO}[\langle\delta{X^2_{LO}}\rangle-
\langle\delta{X^2_{\phi}}\rangle],
\label{cov}
\end{eqnarray}
Here it is assumed that the EN of the two detectors are
not correlated i.e. $\langle\delta{i_{el1}}\delta{i_{el2}}\rangle=\langle\delta{i_{el1}}\rangle\langle\delta{i_{el2}}\rangle$.
As seen from~(\ref{cov}) the covariance coefficient is completely independent of the EN due to the time
averaging of the data and the statistical independence of the noise of
two photodetectors. For the sake of clarity, we assume that the beam of the LO
is shot noise limited i.e. $\langle\delta{X^2_{LO}}\rangle= \langle\delta{X^2_{vac}}\rangle$. The analysis of the formula ~(\ref{cov}) suggests, that the sign of the covariance coefficient is determined by the
statistics of the incoming light. Indeed, if the input state is
squeezed i.e. $\langle\delta{X^2_{\phi}}\rangle<
\langle\delta{X^2_{vac}}\rangle$, the covariance is positive.
In contrast, if the input state is coherent i.e.
$\langle\delta{X^2_{\phi}}\rangle= \langle\delta{X^2_{vac}}\rangle$
the covariance is equal to zero, and finally, if the input
state exhibits classical excess noise i.e.
$\langle\delta{X^2_{\phi}}\rangle> \langle\delta{X^2_{vac}}\rangle$,
the covariance is negative. Therefore, the sign of the
covariance coefficient allows one to carry out a qualitative characterization of the statistics of the measured quadrature. 

The covariance measurements can be also used to estimate the degree of squeezing. However, in calibrating the shot noise level one must resort to the standard HD approach using the methods described above. Once this calibration is obtained for a specific pair of detectors it can be used for any subsequent squeezing measurements exploiting the covariance method. By comparing eq.~(\ref{SNL2}) for the shot noise with eq.~(\ref{cov}) for the covariance we find the following expression for the degree of squeezing
\begin{eqnarray}
\hbox{S}=\langle\delta X_{\phi}^2\rangle = 1-4\frac{\hbox{cov}(i_{1},i_{2})}{V_{SNL}}
\label{squeezingCorr}
\end{eqnarray}    
This expression is seen to be independent on the electronic noise of the detectors, and thus the correlation method yields an electronic noise free estimation of squeezing. The estimation relies, however, on a noise-free calibration of the SNL, that is, $V_{SNL}$ must be measured accurately as mentioned above. 

Let us mention that in the derivation of the covariance coefficient (6), we did not perform a normal ordering procedure of the creation and annihilation operators. On the other hand, the joint detection operator, referred as a \textit{coincidence operator}, which is traditionally introduced in discrete variable quantum optics, is defined through the second normally-ordered intensity moment ~\cite{mandelbook}. In order to find the link between the coincidences operator and the covariance coefficient, we rewrite the normally ordered coincidence operator and apply the commutation relations for $a$ and $a^\dagger$:
\begin{eqnarray}
\langle{:i_{1}i_{2}}:\rangle-\langle{i_{1}}\rangle\langle{i_{2}}\rangle&=&\langle{i_{1}i_{2}}\rangle-\langle{i_{1}}\rangle\langle{i_{2}}\rangle\nonumber\\ \Rightarrow\langle{:i_{1}i_{2}}:\rangle&=&cov(i_{1},i_{2})+\langle{i_{1}}\rangle\langle{i_{2}}\rangle.
\end{eqnarray}
Thus, the covariance coefficient, revealed from the direct product of two photocurrents, corresponds to the normally ordered coincidence operator $\langle{:i_{1}i_{2}}:\rangle$ with subtracted background of the "accidental coincidences", proportional to $
\langle{i_{1}}\rangle\langle{i_{2}}\rangle$. 

It is also useful to see how the covariance coefficient scales with linear losses. Let us first consider a situation where the LO and the input signal experience the
same linear losses $\eta$. From a simple beamsplitter model it
follows that the LO will loose in mean intensity and a portion
of the vacuum fluctuations will be coupled to the input beam. We
obtain the following expression for the attenuated covariance:
\begin{equation}
\hbox{cov}_{\eta}(i_{1},i_{2})=(1-\eta)^2\alpha^2_{LO}[\langle\delta{X^2_{LO}}\rangle-
\langle\delta{X^2_{\phi}}\rangle].
\label{att1}
\end{equation}
This result is confirmed by the general rule, that the second 
normally ordered intensity moment scales with the square of
the transmission, i.e. $\langle{:i_1i_2:}\rangle_\eta=(1-\eta)^2\langle{:i_1i_2:}\rangle$
~\cite{mandelbook}.
We now consider the scenario where the LO power is decreased but without affecting the signal state. In this case the covariance coefficient scales linearly with the transmission, $1-\eta$:
\begin{equation}
\hbox{cov}_{\eta}(i_{1},i_{2})=(1-\eta)\alpha^2_{LO}[\langle\delta{X^2_{LO}}\rangle-
\langle\delta{X^2_{\phi}}\rangle].
\label{att2}
\end{equation}
These linear and quadratic dependencies with respect to the transmission will be experimentally verified in the following sections. 

\section{Experiment}

\begin{figure}
\includegraphics[width=8cm]{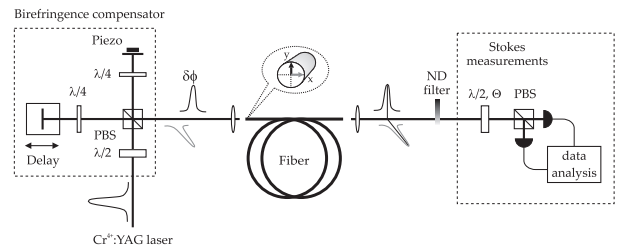}
\caption{The scheme of the polarization squeezing setup. The beam from the Cr$^{4+}$:YAG laser is split on the polarization beam splitter (PBS) after polarization rotation, performed by a half-wave plate (HWP). The birefringence compensator is used to provide a temporal overlap between two orthogonally polarized pulses after propagation through a single mode polarization maintaining fiber. The interference between the pulses is observed by placing a HWP and a PBS in front of two analogue photodetectors. The photocurrents of two detectors are further analyzed by the high-speed digitizer.}
\label{fig2Kerr}
\end{figure}

\begin{figure}
\includegraphics[width=8cm]{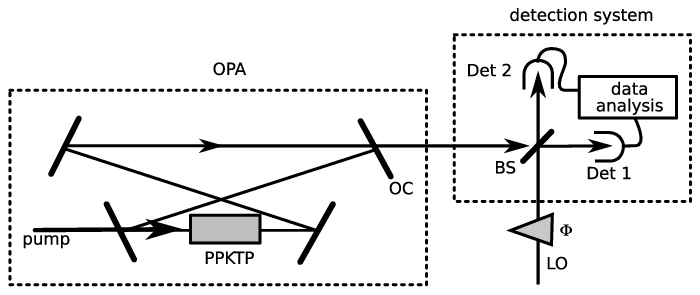}
\caption{The scheme of the OPA based setup. A frequency doubled Nd:YAG laser (fundamental wavelength at 1064 nm) is used to pump the bow type shaped OPA with a 10mm long PPTKP crystal. The squeezed vacuum field, generated by the OPA, is mixed with the fundamental wave of the pump laser on a 50/50 beam splitter (BS), used as a LO, and further detected by two analogue photodetectors (Det1, Det2). The relative phase between the squeezed vacuum field and the LO is controlled by a mirror on a piezo-stage (not shown). The photocurrents of two detectors are further analyzed by the high-speed digitizer.}
\label{fig3OPO}
\end{figure}

The idea described in the previous section was experimentally investigated in ref.~\cite{kriv} for a very bright polarization squeezed state. In that realization, the LO was in the same spatial mode with the squeezed vacuum. Therefore, it was not possible to attenuate the two modes (the squeezed and the LO) independently. In the present paper we extend the previous analysis to include also the investigation of the squeezed vacuum prepared in a mode spatially separated from the LO, thus allowing independent control of the two modes. For completeness, we present in this and the following section both the results of the new experiment as well as the results of ref.~\cite{kriv}.

The setup for the bright squeezing experiment, based on the Kerr nonlinearity in optical fibers, is shown in Fig.~\ref{fig2Kerr}~\cite{polsqueezing, polsqueezing1}. A Cr$^{4+}$:YAG laser generates pulses with a duration of 140 femtoseconds, at a repetition rate of 163 MHz and with a central wavelength of 1497 nm. The laser pulse is first split between two orthogonal polarization modes as well as two different temporal modes using a "birefringence compensator" (see Fig. 2) and secondly launched into a 13.3 meter long single mode polarization maintaining fiber. While propagating through the fiber, the orthogonally polarized pulses experience the Kerr effect, resulting in the squeezing of a certain quadrature in both polarization modes. The temporal overlap between two output pulses is provided by the "birefringence compensator" in front of the fiber as well as the locking loop based on a Stokes parameter measurement of a small portion ($\leq 0.1\%$) of the fibre output. This resulted in a circularly polarized beam at the fiber output, mathematically described by $\left\langle \hat{S}_3\right\rangle\neq0$ and $\left\langle \hat{S}_1\right\rangle=\left\langle \hat{S}_2\right\rangle = 0$ (where $S_1$, $S_2$ and $S_3$ are the three Stokes parameters). The conjugate polarization operators, which can exhibit polarization squeezing, are then found in the so called dark plane given by $\hat{S}_1 - \hat{S}_2$. The polarization squeezing is observed by a Stokes parameter measurement setup with the help of a HWP placed in front of a PBS. Then two analogue high-efficient photodetectors with quantum effiencies $\geq98\%$ are placed in the output ports of the PBS and their photocurrents are analyzed by a high-speed digitizer. Such polarization squeezing is equivalent to a "vacuum" squeezing in a certain polarization mode (here the left circular polarization mode) whereas the orthogonal polarization mode (right circular polarization mode) serves as the intrinsic local oscillator~\cite{polsqueezing,polsqueezing1}. The relative phase between the LO and the squeezed "vacuum" is varied by rotating the HWP placed in front of the PBS in the Stokes measurement setup. Note that since the two polarization modes share the same spatial mode, independent control or attenuation of them is not straightforward. This limitation lead us to also use an alternative source of the squeezed light based on the optical parametric amplifier(OPA) which is described below.

The OPA setup is shown in Fig.~\ref{fig3OPO}. We use a bow tie shaped cavity with a nonlinear crystal, pumped by the second harmonic (532 nm) of a continuous wave single mode field from a Nd:YAG laser. The OPA has a cavity round trip path length of 275mm and the transmittance of the output coupler (OC) is 10\%. A 10mm long periodically poled KTP (PPKTP) crystal is used as a nonlinear medium inside the cavity, and the cavity is phase locked using the  Pound-Drever-Hall locking techniques on counter propagating light field. The squeezed vacuum field at 1064 nm is measured by combining it on 50/50 beam splitter with a LO drawn directly from the laser. The relative phase between the squeezed vacuum and the LO is controlled by a mirror on a piezo
stage. 

In both experimental arrangements we recorded the raw data from
the two detectors by digitizing an AC component of the photocurrents. For this purpose we used a high-speed digitizer with sampling
rates of $2\cdot10^6\hbox{samples/s}$ and $20\cdot10^6\hbox{samples/s}$ and input bandwidths of $150\hbox{KHz}$ and $3\hbox{MHz}$ for the OPA and the Kerr based squeezing setup, respectively. Finally, a simple Matlab script evaluated the variance of the difference current (corresponding to standard HD) and the covariance between the two detector outcomes. This enabled a straightforward comparison between the two methods. 

By changing the phase $\phi$ of the LO (using a half wave plate in Fig.~\ref{fig2Kerr} and mirror on a piezo
stage in Fig.~\ref{fig3OPO}), we were able to choose the
measured quadrature $\delta X_\phi$ to be either squeezed or
anti-squeezed. The measurement of the coherent state was taken
with a shot noise limited beam, whilst the squeezed beam was blocked.

The influence of the EN on the results of the measurements was
studied by attenuating the intensity of the optical
signal using neutral density (ND) filters. In case of strong
attenuation, the produced photocurrent will carry a relatively large contribution from the EN, whereas for no attenuation the EN plays only a minor role. Therefore through attenuation, the transition between EN dominance and the optical noise dominance can be smoothly accessed.    
For the Kerr squeezed state both the LO as well as the squeezed state was attenuated whereas for the OPA experiment only the LO underwent the attenuation.

\section{Results and discussion}

\begin{figure}
\includegraphics[width=7cm]{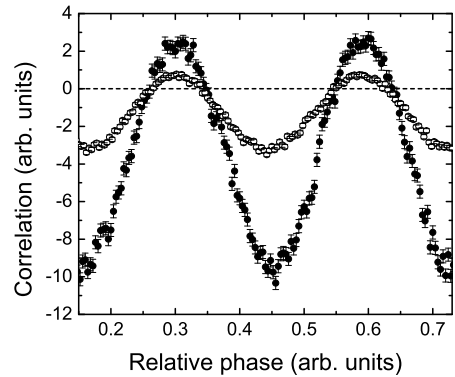}
\caption{The covariance versus the
phase of the LO at two different powers of the LO (open and filled circles) for the OPA based squeezing setup. Positive covariance values correspond to the measurement of the squeezed quadrature, whilst the negative covariance values correspond to the measurement of the antisqueezed
quadrature.}
\label{fig4}
\end{figure}

\begin{figure}
\includegraphics[width=7cm]{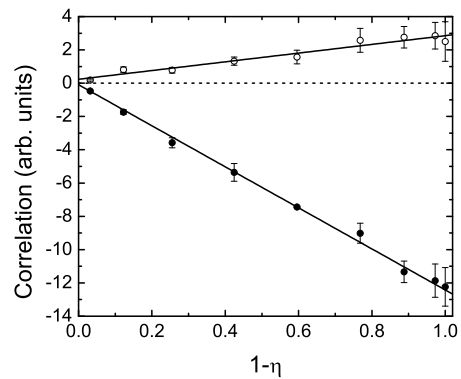}
\caption{The covariance at the maximally squeezed (open circles) and the maximally antisqueezed quadrature (filled circles)  versus the transmittance of the LO (1-$\eta$).}
\label{fig5}
\end{figure}

\begin{figure}
\includegraphics[width=7cm]{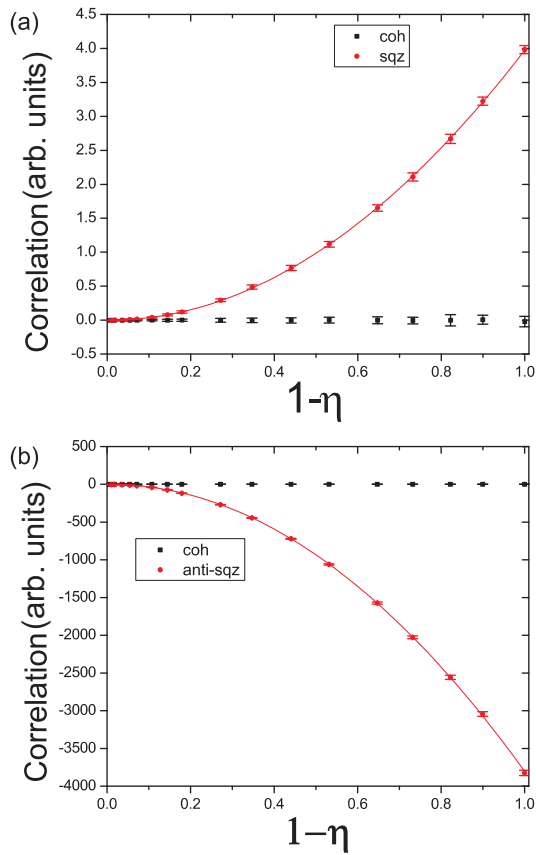}
\caption{Effect of the transmittance of the LO (1-$\eta$) on the covariance at (a) the maximally squeezed and (b) the maximally antisqueezed quadrature. In both plots we compare the covariance of a squeezed beam (red) to a calibration measurement using a coherent beam (black).
}
\label{fig5extra}
\end{figure}

First we investigated the covariance as a function of the phase $\phi$ of the LO. In Fig.~\ref{fig4} we present the results of such measurements using the OPA setup with two different powers of the LO. Positive covariance
corresponds to the measurement of squeezed quadratures, whilst
the negative covariance corresponds to the measurement of the
anti-squeezed quadratures. We see that the phase interval for which squeezing is witnessed is independent on the power of the local oscillator. With other words, despite the presence of a relatively large contribution of electronic noise in the detectors while using a weak LO, the presence of squeezing in the signal is still evident. 
A similar trend of the covariance was observed for the Kerr based squeezing source~\cite{kriv}. 
However, in that case the covariance modulus of the anti-squeezing quadrature were much larger due to the larger impurity of the Kerr squeezed state~\cite{polsqueezing2}.

Correlation plots as the ones in Fig.~\ref{fig4} were made for various different powers of the LO, and for each realization we recorded the maximum correlation and anti-correlation. The results are shown in Fig.~\ref{fig5} where the covariance is plotted against the power of the LO. It clearly demonstrates the linear dependence as predicted by equation~(\ref{att2}). We performed similar measurements for the Kerr squeezed state. This time, however, the squeezed mode as well as the LO mode were attenuated simultaneously, which according to eq.~(\ref{att1}) yields a quadratic dependence in 1-$\eta$. This was experimentally confirmed by the measurements as summarized in Fig.~\ref{fig5extra}. 

In the next step of the experiment we wanted to perform an electronic noise free measurement of the degree of squeezing. For this purpose we first calibrated the shot noise level. It was important that this calibration was not affected by the electronic noise of the detectors as this otherwise would yield a false estimation of the SNL and a corresponding overestimation of the degree of squeezing. The calibration was therefore established at very high powers of the LO, making the first term in eq.~(\ref{SNL}) very large and the following EN terms negligible. Using the fact that the shot noise scales linearly with power, we could easily determine the shot noise at all powers from the high power measurement data. These calibrations (one for the OPA experiment and one for the Kerr experiment) were then subsequently used for establishing the degree of squeezing. 

\begin{figure}
\includegraphics[width=7cm]{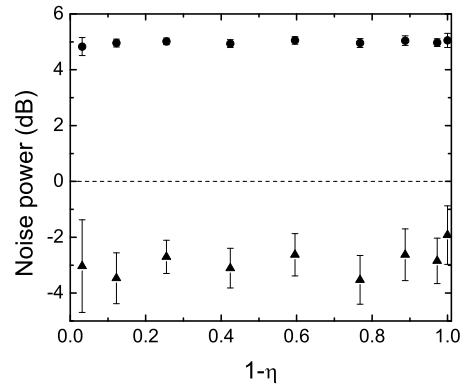}
\caption{The degree of squeezing (triangles) and antisqueezing (circles) revealed from the correlation analysis versus the transmittance of the LO (1-$\eta$) in the OPA based squeezing setup. The relatively large error bars are mostly due to statistical fluctuations.}
\label{fig6}
\end{figure}

\begin{figure}
\includegraphics[width=7cm]{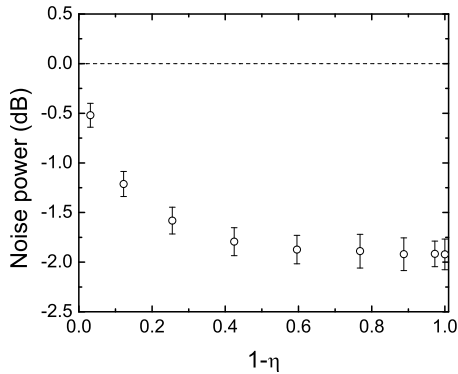}
\caption{Measurement along the squeezed quadrature in the OPA based squeezing setup using homodyne detection. The plot shows, how the degree of squeezing depends on the transmission of the LO (1-$\eta$).}
\label{fig7}
\end{figure}

\begin{figure}
\includegraphics[width=7cm]{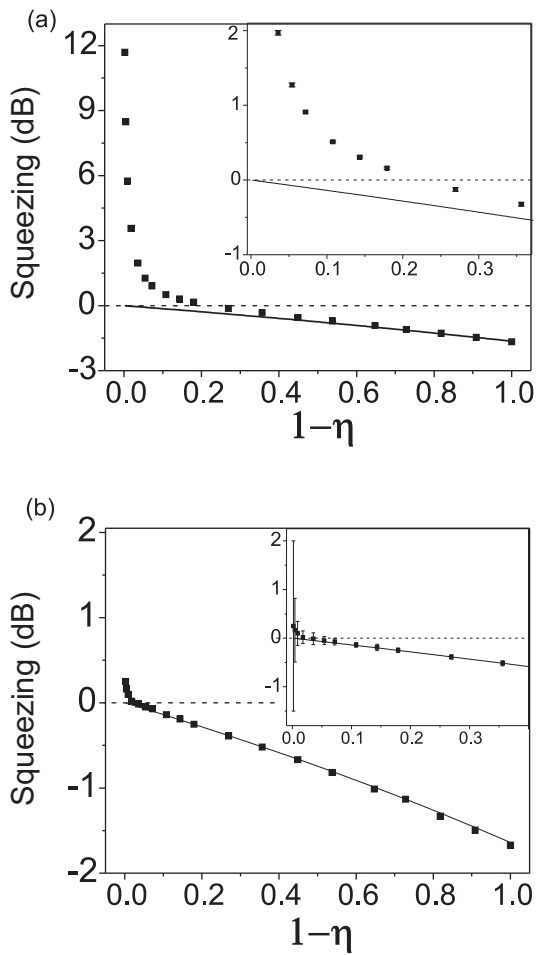}
\caption{Experimental dependence of the amount of squeezing on the transmission (1-$\eta$) (a) for homodyne detection and (b) for the covariance method. The solid lines represent the theoretical
prediction without accounting for the EN. The horizontal dashed line at zero marks the border between squeezing and classical excess noise.}
\label{fig8}
\end{figure}

From the correlation measurements, presented in Fig.~\ref{fig5} and~\ref{fig5extra}, and the corresponding shot noise calibrations, we use eq.~(\ref{squeezingCorr}) to compute the degree of squeezing. The results are summarized in Fig.~\ref{fig6} for the OPA squeezer and Fig.~\ref{fig8} for the Kerr squeezer. We clearly see from Fig.~\ref{fig6} that the measured degree of squeezing is almost constant as the LO power is decreased although the relative contribution from the EN becomes large. This should be compared with traditional homodyne detection of the degree of squeezing, the measurements of which are shown in Fig.~\ref{fig7}. Here we see the expected trend for homodyne detection, that the measured degree of squeezing goes down as the relative EN increases. This is indeed the main conclusion of this paper; by using the correlation method, it is possible to reliably estimate the degree of squeezing independent of the EN (even though the LO is very dim) in contrast to the standard homodyne measurement approach which is affected by electronic noise. This also holds true for a LO free setup where a very dim squeezed beam is directly measured (using a beam splitter and two detectors). In this case, however, the signs of the correlations will swap and the squeezing will be computable from the expression $S=\frac{4cov(i_1,i_2)}{V_{SNL}}-1$~\cite{leuchs}.

Similar investigations were carried out for the Kerr squeezer as described in ref.\cite{kriv} and summarized in Fig.~\ref{fig8}; in Fig.~\ref{fig8}A and B we plot the measured degree of squeezing versus the transmission using homodyne detection and correlation measurements, respectively. The insets show magnified parts of the plots. Similar conclusions can be drawn from these plots as from Figs.~\ref{fig6} and~\ref{fig7}: The correlation method is capable of measuring squeezing at lower LO powers than for the homodyne detectors. In these experiments, however, the attenuation was applied collectively on both modes which means that the actual degree of squeezing was degraded as the transmittivity goes down. 

We should emphasize that the "removal" of electronic noise is a result of the averaging procedure after detection as mentioned in section II. Every single measurement data is largely affected by electronic noise which means that the correlation method is not yielding electronic noise free data "in real time" but only after several measurements of an ensemble and subsequently averaging. Moreover, the single measurement does not give information about the amplitude of the signal and thus cannot in general be used in feedforward based quantum information protocols.

\section{Conclusions}

In conclusion, we have studied the implementation of a correlation measurement technique on two different sources of squeezed light generated from an optical parametric oscillator and from an optical fiber. We showed that this method can be used to estimate the degree of squeezing independent of the amount of electronic noise in the detectors provided that their noises are uncorrelated. This is in stark contrast to using the standard homodyne detection to estimate the degree of squeezing, where electronic noise might affect the measurements, especially when the LO power is low.

The correlation measurement strategy is particular useful for measuring the degree of squeezing of a squeezed light that is neither matchable to any local oscillator modes nor has a high coherent excitation. For such cases, the electronic noise becomes a critical issue which can be overcome by the correlation measurement method.

We appreciate discussions with M. Chekhova, R. Filip, and Ch.
Marquardt. This work was supported by the EU project COMPAS and the
Danish Research Council (FTP). One of us (L.A.K.) acknowledges the
support of the Alexander von Humboldt foundation.

\end{document}